\documentclass[12pt]{article}
\usepackage{amssymb}
\usepackage{amsmath}
\usepackage{amstext}
\usepackage{graphicx,epsfig}
\usepackage{epsfig}
\usepackage{color}
\usepackage{parskip}
\usepackage{cite}
\usepackage{tikz}
\usepackage{mathtools}
\usepackage[utf8]{inputenc}

\usetikzlibrary{positioning, trees,decorations, decorations.markings, decorations.pathmorphing, arrows, shapes.geometric, snakes}

\allowdisplaybreaks

\newcommand{\Comment}[1]{{}}
\definecolor{MyDarkBlue}{rgb}{0.15,0.15,0.45}
\usepackage[linktocpage=true]{hyperref}
\hypersetup{
colorlinks=true,
citecolor=MyDarkBlue,
linkcolor=MyDarkBlue,
urlcolor=MyDarkBlue,
pdfauthor={Kurt Hinterbichler},
pdftitle={},
pdfsubject={hep-th}
}

\newcommand{\be}{\begin{equation}}
\newcommand{\ee}{\end{equation}}
\newcommand{\bal}{\begin{align}}
\newcommand{\eal}{\end{align}}
\newcommand{\bals}{\begin{align*}}
\newcommand{\eals}{\end{align*}}
\newcommand{\bea}{\begin{eqnarray}}
\newcommand{\eea}{\end{eqnarray}}
\newcommand{\beas}{\begin{eqnarray*}}
\newcommand{\eeas}{\end{eqnarray*}}

\def\({\left(}
\def\){\right)}

\numberwithin{equation}{section}

\begin{document}

\tikzset{
    photon/.style={decorate, decoration={snake}, draw=red},
    graviton/.style={decorate, decoration={snake}, draw=black},
  mgraviton/.style={draw=black},
    electron/.style={draw=blue, postaction={decorate},
        decoration={markings,mark=at position .55 with {\arrow[draw=blue]{>}}}},
    gluon/.style={decorate, draw=magenta,
        decoration={coil,amplitude=4pt, segment length=5pt}} 
}

\begin{center}
{\LARGE {A Universal Bound on the Strong Coupling Scale of a Gravitationally Coupled Massive Spin-2 Particle\\ \vspace{.2cm} }}
\end{center}
\vspace{2truecm}
\thispagestyle{empty} \centerline{
{\large {James Bonifacio,}}$^{}$\footnote{E-mail: \Comment{\href{mailto:jjb239@case.edu}}{\tt james.bonifacio@case.edu}}
{\large { Kurt Hinterbichler$^{}$}}\footnote{E-mail: \Comment{\href{mailto:kurt.hinterbichler@case.edu}}{\tt kurt.hinterbichler@case.edu}}
}

\vspace{1cm}

\centerline{\it ${}^{\rm }$CERCA, Department of Physics, Case Western Reserve University, }
\centerline{\it 10900 Euclid Ave, Cleveland, OH 44106, USA}

\begin{abstract}

We find a model-independent upper bound on the strong coupling scale for a massive spin-2 particle coupled to Einstein gravity. Our approach is to directly construct tree-level scattering amplitudes for these degrees of freedom and use them to find the maximum scale of perturbative unitarity violation. The highest scale is $\Lambda_3=\left(m^2M_P\right)^{1/3}$, which is saturated by ghost-free bigravity. The strong coupling scale can be further raised to $M_P$ if the kinetic term for one particle has the wrong sign, which uniquely gives the amplitudes of quadratic curvature gravity. We also discuss the generalization to massive higher-spin particles coupled to gravity.

\end{abstract}


\newpage

\setcounter{tocdepth}{3}
\tableofcontents

\section{Introduction}
\parskip=5pt
\normalsize

Every particle that interacts with Einstein gravity in flat spacetime must do so through a minimal coupling vertex with a universal gravitational strength~\cite{Weinberg:1964ew}. This property results in powerful constraints on the allowed particles and interactions that can exist alongside Einstein gravity.  One notable such constraint is that in flat spacetime there can be no local theories of \textit{massless} higher-spin particles interacting with anything that interacts with Einstein gravity---the required gravitational minimal coupling interactions are incompatible with higher-spin gauge invariance \cite{Aragone:1979hx, Metsaev:2005ar, Boulanger:2008tg,Porrati:2008rm}.  When it comes to \textit{massive} particles, gauge invariance is not a constraint and no longer forbids higher spins from coupling universally to Einstein gravity.  Indeed, such particles exist in nature as unstable hadrons and glueballs.  However, in this case gravity does still impose restrictions, since an isolated massive higher-spin particle is not expected to remain a fundamental point-like particle up to the Planck scale~\cite{Cucchieri:1994tx, Arkani-Hamed:2017jhn}.  Due to the mandatory gravitational interactions that are required by the equivalence principle, there should exist a maximum strong coupling scale beyond which the local effective field theory (EFT) of an isolated massive higher-spin particle must break down and this scale goes to zero with the mass of the particle.

First let us review what happens for low-spin particles. Consider a spin-0 field $\phi$ of mass $m$ minimally coupled to gravity, which is described by the action
\be
S = \frac{1}{2} \int d^4 x \sqrt{-g} \left(M_P^2 R-\partial_{\mu} \phi \partial^{\mu} \phi -m^2 \phi^2 \right),
\ee
where $M_P$ is the reduced Planck mass.
By expanding the metric around a flat background $\eta_{\mu\nu}$, we expose the interactions between the scalar and the canonically normalized massless spin-2 graviton described by $h_{\mu \nu}$, where $g_{\mu \nu} = \eta_{\mu \nu} + 2 h_{\mu \nu}/M_P$.
These interactions result in a nonzero amplitude for tree level scalar-scalar scattering via graviton exchange.  This amplitude is given by
\be
\mathcal{A} = \frac{1}{4M_P^2} \frac{\left(s^2+t^2+u^2\right)^2}{s t u}+{\cal O}\(m^2\)\, ,
\ee
where $s$, $t$, $u$ are the usual Mandelstam variables (defined in \eqref{mandelstamdefe} below) and we have omitted terms that are subleading for energies much larger than $m$.
This amplitude is small for low-energy, large-angle scattering, but it violates perturbative unitarity bounds for energies around the Planck scale---the scale where the local EFT of gravity breaks down~\cite{Han:2004wt}. We can thus safely assume that these high-energy violations of perturbative unitarity are resolved by the same unknown theory of quantum gravity that tames the growth and divergences of graviton scattering amplitudes. A similar story holds for minimally coupled massive spin-$1/2$ and spin-1 particles. 

For massive particles with spin $3/2$ and above, the story is qualitatively different. A minimally coupled massive higher-spin particle will scatter via graviton exchange to give an amplitude that grows like some high power of energy suppressed by an energy scale
\be \label{eq:cutoff}
\Lambda_k \equiv \left( m^{k-1} M_P \right)^{1/k},
\ee
where $k>1$ is some number and $m$ is the mass of the higher-spin particle.  (We must assume that $m < M_P$, otherwise we cannot sensibly talk about the massive particle in an EFT with gravity). At the scale $\Lambda_k$, tree amplitudes become of order one and violate partial-wave unitarity bounds.\footnote{Throughout we consider only the parametric dependence of the strong coupling scale on the masses and couplings. Fixing the precise coefficients requires decomposing amplitudes into spinning partial-waves and imposing unitarity.} Unlike for spins less than $3/2$, this scale is parametrically smaller than $M_P$ and vanishes in the massless limit.  There is some intrinsic limit to the perturbative validity of the local EFT for a massive higher-spin particle, set by the scale \eqref{eq:cutoff}, and beyond this limit new physics or strong coupling effects must become important.  This implies that massive higher-spin states cannot exist as isolated elementary particles in flat spacetime all the way up to the Planck scale---they are always accompanied by other particles or strong coupling effects that come in at a lower scale, as in perturbative string theory and confining gauge theories. 

We can understand this quantitatively by determining the maximum $\Lambda_k$ for a given spectrum of particles.  For a single massive spin-$s$ particle, naive power counting from minimal coupling implies that scattering the scalar longitudinal mode of the particle via graviton exchange generically gives $k= 2s+1$ in \eqref{eq:cutoff}.  However, it is possible that non-minimal couplings or other local terms\footnote{We assume that the number of derivatives is arbitrary but finite, see Ref.~\cite{Bonifacio:2018vzv} for more discussion of this point.} can be added to the Lagrangian to raise the strong coupling scale, so determining the maximum possible scale is a non-trivial problem.  Here we will study this problem for the case of a massive spin-2 particle coupled to gravity. We will show that the maximum strong coupling scale in a theory with these degrees of freedom is $\Lambda_3 = \left(m^2 M_P \right)^{1/3}$, and that this scale can be further raised to $M_P$ if wrong-sign kinetic terms are permitted. Our approach is to directly construct the four-point tree amplitude with external massive spin-2 states of any polarization, generalizing the analysis of Ref.~\cite{Bonifacio:2018vzv} to include a massless spin-2 particle.  We then look for the amplitudes that maximize the strong coupling scale among those containing the requisite minimal coupling interactions.  As discussed in Ref.~\cite{Bonifacio:2018vzv}, this circumvents the problem of having to consider Lagrangians with arbitrary numbers of derivatives.

Our results are in the same spirit as Refs.~\cite{Porrati:2008an,Porrati:2008gv,Porrati:2008ha}, where the maximum strong coupling scale was found for massive spinning particles charged under electromagnetism. This bound was obtained by using the St\"uckelberg formalism and cohomological methods to look for Lagrangian interactions that could not be removed by field redefinitions or by adding additional local operators. For a massive spin-$s$ particle with charge $q$, it was found that the maximum strong coupling scale is
\be \label{eq:EMcutoffs}
\Lambda_{\rm } = \frac{m}{q^{1/({2s-1)}}}.
\ee
This result assumes the presence of electromagnetic minimal coupling interactions, although these are not compulsory, unlike gravitational minimal coupling interactions. This method has also been applied to the case of a gravitationally coupled massive spin-$3/2$ particle in Ref.~\cite{Rahman:2011ik}, where it was found that the maximum strong coupling scale is given by\footnote{An example of a theory realizing this strong coupling scale is $\mathcal{N}=1$ broken supergravity with the scalar and pseudoscalar from the chiral supermultiplet integrated out~\cite{Casalbuoni:1988sx}.} $\Lambda_2 \equiv (m M_P)^{1/2}$. It was conjectured in Ref.~\cite{Rahman:2009zz} that the maximum strong coupling scale for a massive spin-$s$ particle coupled to gravity is given by 
\be 
\Lambda_{2s-1}=\left(m^{2s-2} M_P \right)^{1/(2s-1)}.
\ee
This conjecture agrees with the results of this paper for $s=2$. We will provide some additional motivation for the spin-$s$ case in Section~\ref{sec:discussion} using the result \eqref{eq:EMcutoffs} for charged particles and the weak gravity conjecture.  Other related works include the study of massive higher-spin particles propagating in gravitational backgrounds \cite{Deser:1983mm,Higuchi:1986py, Bengtsson:1994vn,Buchbinder:1999ar,Buchbinder:1999be,Buchbinder:2000fy,Porrati:2000cp,Zinoviev:2006im,Buchbinder:2012xa,Zinoviev:2013hac,Bernard:2014bfa,Bernard:2015mkk,Bernard:2015uic,Bernard:2017tcg,Cortese:2013lda,Rahman:2016tqc,Kulaxizi:2014yxa}.

The outline of this paper is as follows: in Section~\ref{sec:vertices} we construct all the necessary on-shell cubic and quartic vertices for an EFT of a massive spin-2 particle coupled to gravity. In Section~\ref{sec:bound} we review how to use these vertices to calculate four-point tree amplitudes with a given strong coupling scale.  In Section~\ref{sec:results} we find the highest possible strong coupling scale in a theory with these degrees of freedom and discuss several examples subject to our results. We conclude in Section~\ref{sec:discussion} by discussing some implications of our results and the generalization to higher spins. In Appendix~\ref{app:gauge} we discuss constraints on cubic vertices from gauge invariance. 

\bigskip
\noindent
{\bf Conventions}:
We work in flat four-dimensional spacetime and use the mostly plus metric signature convention, $\eta_{\mu\nu}=(-,+,+,+)$.  We (anti)-symmetrize indices with weight one.

\section{On-shell vertices} \label{sec:vertices}

\begin{figure}[!ht]
\begin{center}
\begin{tikzpicture}[node distance=.8cm and 1.2cm]
\coordinate (s1);
\coordinate[below right=of s1] (svertex1);
\coordinate[below left=of svertex1] (s2);
\coordinate[right=1.2cm of svertex1] (svertex2);
\coordinate[above right=of svertex2] (s3);
\coordinate[below right=of svertex2] (s4);

\draw[mgraviton] (s1) -- (svertex1);
\draw[mgraviton] (svertex1) -- (s2);
\draw[mgraviton] (s3) -- (svertex2);
\draw[mgraviton] (svertex2) -- (s4);
\draw[mgraviton] (svertex1) -- (svertex2);

\coordinate[right=2cm of svertex2,label=left:$+$] (mid1);
\coordinate[right=1.2cm of mid1] (tmid);

\coordinate[above=.5cm of tmid] (tvertex1);
\coordinate[above left=of tvertex1] (t1);
\coordinate[above right=of tvertex1] (t3);
\coordinate[below=1cm of tvertex1] (tvertex2);
\coordinate[below left=of tvertex2] (t2);
\coordinate[below right=of tvertex2] (t4);

\draw[mgraviton] (t1) -- (tvertex1);
\draw[mgraviton] (t3) -- (tvertex1);
\draw[mgraviton] (tvertex1) -- (tvertex2);
\draw[mgraviton] (t2) -- (tvertex2);
\draw[mgraviton] (t4) -- (tvertex2);

\coordinate[right=3cm of mid1,label=left:$+$] (mid2);

\coordinate[right=3.5cm of tvertex1] (uvertex1);
\coordinate[above left=of uvertex1] (u1);
\coordinate[above right=of uvertex1] (u3);
\coordinate[below=1cm of uvertex1] (uvertex2);
\coordinate[below left=of uvertex2] (u2);
\coordinate[below right=of uvertex2] (u4);

\draw[mgraviton] (u1) -- (uvertex1);
\draw[mgraviton] (u3) -- (uvertex2);
\draw[mgraviton] (uvertex1) -- (uvertex2);
\draw [white, line width=6] (9.365,-.9) -- (9.5,-.7);
\draw[mgraviton] (u2) -- (uvertex2);
\draw[mgraviton] (u4) -- (uvertex1);

\coordinate[below=3cm of s1] (s1b);
\coordinate[below right=of s1b] (svertex1b);
\coordinate[below left=of svertex1b] (s2b);
\coordinate[right=1.2cm of svertex1b] (svertex2b);
\coordinate[above right=of svertex2b] (s3b);
\coordinate[below right=of svertex2b] (s4b);

\coordinate[left=1.6cm of svertex1b,label=left:$+$];

\draw[mgraviton] (s1b) -- (svertex1b);
\draw[mgraviton] (svertex1b) -- (s2b);
\draw[mgraviton] (s3b) -- (svertex2b);
\draw[mgraviton] (svertex2b) -- (s4b);
\draw[graviton] (svertex1b) -- (svertex2b);

\coordinate[right=2cm of svertex2b,label=left:$+$] (mid1b);
\coordinate[right=1.2cm of mid1b] (tmidb);

\coordinate[above=.5cm of tmidb] (tvertex1b);
\coordinate[above left=of tvertex1b] (t1b);
\coordinate[above right=of tvertex1b] (t3b);
\coordinate[below=1cm of tvertex1b] (tvertex2b);
\coordinate[below left=of tvertex2b] (t2b);
\coordinate[below right=of tvertex2b] (t4b);

\draw[mgraviton] (t1b) -- (tvertex1b);
\draw[mgraviton] (t3b) -- (tvertex1b);
\draw[graviton] (tvertex1b) -- (tvertex2b);
\draw[mgraviton] (t2b) -- (tvertex2b);
\draw[mgraviton] (t4b) -- (tvertex2b);

\coordinate[right=3cm of mid1b,label=left:$+$] (mid2b);

\coordinate[right=3.5cm of tvertex1b] (uvertex1b);
\coordinate[above left=of uvertex1b] (u1b);
\coordinate[above right=of uvertex1b] (u3b);
\coordinate[below=1cm of uvertex1b] (uvertex2b);
\coordinate[below left=of uvertex2b] (u2b);
\coordinate[below right=of uvertex2b] (u4b);

\draw[mgraviton] (u1b) -- (uvertex1b);
\draw[mgraviton] (u3b) -- (uvertex2b);
\draw[graviton] (uvertex1b) -- (uvertex2b);
\draw [white,  line width=6] (9.365,-3.9) -- (9.5,-3.7);
\draw[mgraviton] (u2b) -- (uvertex2b);
\draw[mgraviton] (u4b) -- (uvertex1b);

\end{tikzpicture}
\end{center}
\caption{\small Exchange diagrams for the scattering of a massive spin-2 particle by exchanging a massive spin-2 particle or a graviton.  Solid lines denote the massive spin-2 particle and wavy lines denote the massless spin-2 graviton.}
\label{fig:exchange}
\end{figure}
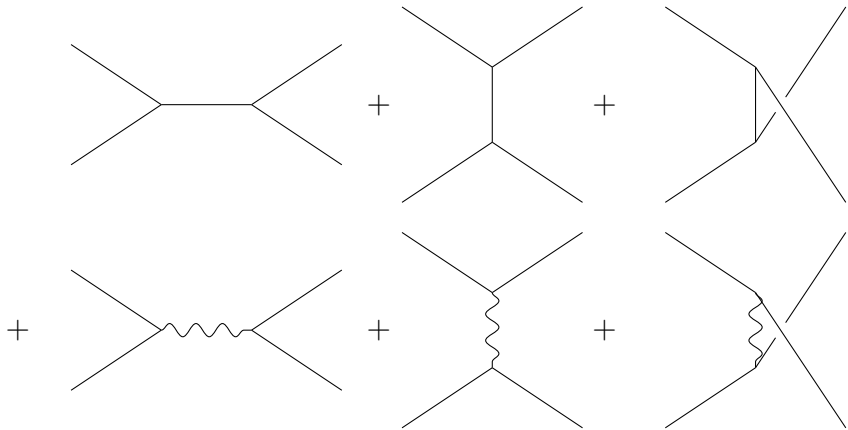

Our strategy to find the maximum strong coupling scale is to construct the most general scattering amplitude allowed by basic principles such as Lorentz invariance, locality, and unitarity.  We will find that the maximum strong coupling scale for a massive spin-2 particle coupled to gravity can be determined from the amplitude with four external massive spin-2 particles.  We do not need to consider other external particle configurations because there are examples of EFTs that saturate the scale we find from the amplitude with four massive external particles.  The contributing diagrams are the exchange diagrams shown in Figure \ref{fig:exchange}, and the contact diagram shown in Figure \ref{fig:4pt}.  
In this section we list all the on-shell vertices that we need to construct these diagrams.

We classify vertices following the approach of Refs.~\cite{Costa:2011mg,Kravchuk:2016qvl}. Similar constructions for spin-2 vertices can be found in Refs.~\cite{Bonifacio:2017nnt, Bonifacio:2018vzv}.  For simplicity, we will assume that parity is conserved.  Let $i,j,\ldots$ index the various spin-2 particles with masses $m_i$ in some vertex.  Each particle has a polarization tensor $\epsilon^i_{\mu \nu}$ that is symmetric, transverse, and traceless.  We write these in terms of auxiliary vectors $z^i_{\mu}$, such that  $\epsilon^i_{\mu \nu} = z^i_{\mu} z^i_{\nu}$. Local parity-even vertices are built from the Lorentz invariant contractions of $z^i_{\mu}$ and the momenta $p^i_{\mu}$, which in this section we take to be incoming. These contractions are denoted by
\be p_{ij}\equiv p^{i}_{\mu} p^{j, \mu}, \quad z_{ij}\equiv z^{i}_{\mu} z^{j, \mu} , \quad zp_{ij}\equiv z^{i}_{\mu} p^{j, \mu}.\ee 
They satisfy $z_{ij} = z_{ji}$, $z_{ii}= 0$, and $zp_{ii}=0$, reflecting the symmetry, tracelessness, and transversality of the polarization tensors, respectively.  We also have $p_{ij}=p_{ji}$, $p_{ii}=-m_i^2$, and $\sum_i p^i_{\mu}=0$.

\subsection{Cubic vertices\label{cubicsubsec}}

\begin{figure}[!ht]
\begin{center}
\begin{tikzpicture}[node distance=1cm and 1.5cm]
\coordinate (vertex1);
\coordinate[above left=of vertex1, label=left:$1$] (e1);
\coordinate[below left=of vertex1, label=left:$2$] (e2);
\coordinate[right=1.50cm of vertex1, label=right:$3$] (e3);

\draw[mgraviton] (e1) -- (vertex1);
\draw[mgraviton] (e2) -- (vertex1);
\draw[mgraviton] (e3) -- (vertex1);

\coordinate[right= 5.5cm of vertex1] (vertex2);
\coordinate[above left=of vertex2, label=left:$1$] (e1);
\coordinate[below left=of vertex2, label=left:$2$] (e2);
\coordinate[right=1.50cm of vertex2, label=right:$3$] (e3);

\draw[mgraviton] (e1) -- (vertex2);
\draw[mgraviton] (e2) -- (vertex2);
\draw[graviton] (e3) -- (vertex2);
\end{tikzpicture}
\end{center}
\caption{\small The three-point vertices needed for the massive four-point amplitude.}
\label{fig:cubics}
 \end{figure}
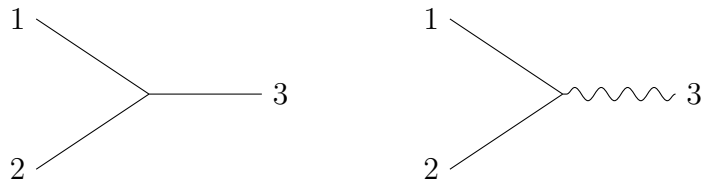

The on-shell cubic vertices for spin-$2$ particles are made from sums of building blocks of the form
\be \label{eq:cubic}
z_{12}^{n_{12}} z_{13}^{n_{13}} z_{23}^{n_{23}} zp_{12}^{m_{12}} zp_{23}^{m_{23}} zp_{31}^{m_{31}} ,
\ee
where the exponents $n_{ij}$ and $m_{ij}$ are non-negative integers that satisfy
\begin{subequations}
\begin{align}
n_{12}+n_{13}+m_{12}& =2, \\
n_{12}+n_{23}+m_{23}& =2, \\
n_{13}+n_{23}+m_{31}& = 2.
\end{align}
\end{subequations}
There are 11 solutions to these equations.  The cubic vertices which appear in Figure \ref{fig:exchange} are those with three identical massive particles and those with two identical massive particles and a massless particle, as depicted in Figure~\ref{fig:cubics}. To find these vertices, we need to look for combinations of the building blocks \eqref{eq:cubic} that are invariant under interchanging the identical particles. Additionally, if particle $i$ is massless we need the vertex to be gauge invariant, which corresponds to invariance under $z_i \rightarrow z_i + \xi p_i$ to first order in $\xi$.

\subsubsection{Three massive particles}

First we consider the vertex containing three identical massive spin-2 particles. There are five possible structures and they are given by
\begin{subequations}
\begin{align}
\mathcal{A}_1 = & z_{12} z_{13} z_{23},  \\
\mathcal{A}_2 = & z_{23}^2 zp_{12}^2+z_{13}^2 zp_{23}^2+z_{12}^2 zp_{31}^2, \label{eq:spin2cubicGR}\\
\mathcal{A}_3 = & z_{13} z_{23}zp_{12}zp_{23}+z_{12}z_{23}zp_{12}zp_{31}+z_{12}z_{13}zp_{23} zp_{31}, \\
\mathcal{A}_4 = &zp_{12} zp_{23} zp_{31} \left( z_{12}zp_{31}+z_{23} zp_{12} +z_{13} zp_{23} \right),\\
\mathcal{A}_5 = & zp_{12}^2 zp_{23}^2 zp_{31}^2.
\end{align}
\end{subequations}
These are invariant under all permutations of the particles.
In four dimensions they are related by a dimensionally-dependent identity,
\be
4 \mathcal{A}_4 - 2 m^2 \left( \mathcal{A}_2+ \mathcal{A}_3 \right) +3m^4 \mathcal{A}_1=0.
\ee
The general cubic vertex in four dimensions can thus be parametrized by
\be
\mathcal{V}_a = i \left( a_1\mathcal{A}_1+a_2\mathcal{A}_2+a_3\mathcal{A}_3+a_5\mathcal{A}_5 \right), \label{cubicseqpa}
\ee
where $a_i$ are real cubic couplings. This vertex is the correct one for a particle that transforms as a tensor under parity. The massive particle could also transform under parity as a pseudotensor, and this would be consistent with our assumption of parity conservation. We discuss this case in Section~\ref{sec:results}.

\subsubsection{Two massive particles and one massless particle}
Now we consider the vertex containing one graviton and two identical massive spin-2 particles. There are six possible structures and they are given by
\begin{subequations}
\begin{align}
\mathcal{B}_1 &= z_{12}^2 zp_{31}^2, \\
\mathcal{B}_2 &= z_{12} zp_{31}\left(z_{23} zp_{12} + z_{1 3} zp_{2 3}\right),  \\
\mathcal{B}_3 &= (z_{2 3} zp_{1 2}+ z_{1 3} zp_{2 3})^2, \\
\mathcal{B}_4 &= zp_{12} zp_{2 3} zp_{3 1}\left(z_{2 3} zp_{1 2} + z_{13} zp_{2 3}\right) ,\\
\mathcal{B}_5 &=  z_{1 2} zp_{1 2} zp_{2 3} zp_{31}^2, \\
\mathcal{B}_6 &=   zp_{12}^2 zp_{23}^2 zp_{31}^2,
\end{align}
\end{subequations}
where we have taken particle three to be the massless one. These structures have a symmetry under interchanging particles one and two, and they are invariant under the gauge transformation for particle three,
\be
z_3 \rightarrow z_3 + \xi p_3,
\ee
to first order in $\xi$. In four dimensions these satisfy the identity
\be
2 \mathcal{B}_5+2\mathcal{B}_4-m^2\mathcal{B}_3=0.
\ee
The general cubic vertex in four dimensions can thus be parametrized by
\be \label{eq:bvertex}
\mathcal{V}_b = i \left( b_1\mathcal{B}_1+b_2\mathcal{B}_2+b_3\mathcal{B}_3+b_4\mathcal{B}_4+b_6\mathcal{B}_6 \right), 
\ee
where $b_i$ are real cubic couplings.

\subsubsection{Three massless particles}

There will also be cubic vertices describing graviton self-interactions. We will assume to start with that these interactions include those described by general relativity. This follows from the assumption that the spin-2 gauge symmetry is nonlinear~\cite{Wald:1986bj}, or that the gravitational force is long range. The massless spin-2 cubic vertex thus includes
\be \label{eq:GRvertex}
\mathcal{V}_{\rm GR} = i\frac{2}{M_P} \left(z_{23} zp_{12}+z_{13} zp_{23}+z_{12} zp_{31}\right)^2.
\ee
This vertex is not needed directly for the four-point amplitude with only massive external particles. However, the existence of this vertex does affect the allowed values of the cubic couplings in the vertex \eqref{eq:bvertex}, since higher-point amplitudes with external gravitons must be gauge invariant. In particular, the cubic couplings must satisfy
\be \label{eq:bconstraints}
2 b_1 = b_2 =  \frac{4}{M_P},
\ee
as discussed further in Appendix~\ref{app:gauge}.
These values of the couplings are required in any theory of a massive and massless spin-2 particle which contains the Einstein-Hilbert vertex \eqref{eq:GRvertex}.

\subsection{Quartic vertices}

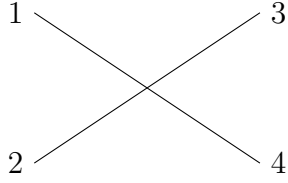
\begin{figure}[!ht]
\begin{center}
\begin{tikzpicture}[node distance=1cm and 1.5cm]
\coordinate[label=left:$1$] (e1);
\coordinate[below right=of e1] (vertex);
\coordinate[below left=of vertex, label=left:$2$] (e2);
\coordinate[above right=of vertex, label=right:$3$] (e3);
\coordinate[below right=of vertex, label=right:$4$] (e4);

\draw[mgraviton] (e1) -- (vertex);
\draw[mgraviton] (e2) -- (vertex);
\draw[mgraviton] (e3) -- (vertex);
\draw[mgraviton] (e4) -- (vertex);
\end{tikzpicture}
\end{center}
\caption{ \small The four-point contact vertex with external massive spin-2 particles.}
\label{fig:4pt}
 \end{figure}
 
 We also need the quartic contact vertices containing four massive spin-2 particles, as depicted in Figure~\ref{fig:4pt}.  We can write the most general such quartic vertex as
\be \label{eq:contactamp}
\mathcal{V}_{\rm contact} = \sum_{I =1}^{201} f_{I }(s,t) \mathbb{T}_{I }(z, p),
\ee
where the 201 tensor structures $\mathbb{T}_{I }(z,p)$ encode the different ways of contracting the polarization tensors and $f_{I }(s,t)$ are polynomials in the Mandelstam variables~\cite{Costa:2011mg, Kravchuk:2016qvl}. The tensor structures are chosen to be invariant under the permutations of the external states that preserve the Mandelstam variables. These `kinematic' permutations are given by a $\mathbb{Z}_2^2$ subgroup of $S_4$~\cite{Kravchuk:2016qvl}, 
\be \label{eq:kinperms}
\Pi^{\rm kin} = \{ \mathcal{I}, (12)(34), (13)(24), (14)(23) \},
\ee
where $\mathcal{I}$ is the identity element. 
Explicitly, the tensor structures $\mathbb{T}_{I }(z,p)$ are given by the on-shell linearly independent terms of the form
\be
\sum_{\pi_k \in \Pi^{\rm kin}} \pi_k \left( z_{12}^{n_{12}} z_{13}^{n_{13}} z_{14}^{n_{14}} z_{23}^{n_{23}} z_{24}^{n_{24}} z_{34}^{n_{34}} zp_{13}^{m_{13}} zp_{14}^{m_{14}} zp_{21}^{m_{21}} zp_{24}^{m_{24}} zp_{31}^{m_{31}} zp_{32}^{m_{32}} zp_{42}^{m_{42}} zp_{43}^{m_{43}} \right),
\ee
where $\pi_k$ acts by permuting the particle labels and $n_{ij}$ and $m_{ij}$ are non-negative integers that satisfy
\begin{subequations} \label{eq:halfspace}
\begin{align}
n_{12}+n_{13}+n_{14}+m_{13}+m_{14} & =2, \\
n_{12}+n_{23}+n_{24}+m_{21}+m_{24} & =2, \\
n_{13}+n_{23}+n_{34}+m_{31}+m_{32} & =2, \\
n_{14}+n_{24}+n_{34}+m_{42}+m_{43} & =2.
\end{align}
\end{subequations}
By construction, the quartic vertex~\eqref{eq:contactamp} is invariant under $\Pi^{\rm kin}$. However, full Bose symmetry for four identical external particles requires invariance under all of $S_4$, which includes permutations that interchange the Mandelstam variables. We impose these additional permutation symmetries on the contact amplitudes by enforcing crossing constraints. 

Not all of the 201 tensor structures $\mathbb{T}_{I }(z, p)$ are independent in four dimensions due to dimensionally-dependent identities. These identities can be used to write certain tensor structures in terms of others. We must therefore be careful when enforcing permutation constraints on tensor expressions in four dimensions, since such constraints only need to be satisfied up to these identities. Our strategy is to avoid this problem by building an ansatz for the contact amplitude directly out of four-dimensional kinematic variables. We then constrain this ansatz by comparing it to the four-dimensional amplitude produced by the tensor expression~\eqref{eq:contactamp}, as discussed further in the next section. 

\section{Finding the maximum strong coupling scale} \label{sec:bound}

In this section we first discuss kinematics and then review the procedure of Ref.~\cite{Bonifacio:2018vzv} for constructing amplitudes and finding the maximum strong coupling scale. 

\subsection{Kinematics} 
First we specify our four-point scattering kinematics in four dimensions. We work in the center-of-mass frame with the momenta for particle $j$ given by
\be
p^j_{ \mu} = \left( E, p \sin \theta_j, 0, p \cos \theta_j \right),
\ee
where $\theta_1=0$, $\theta_2=\pi$, $\theta_3 = \theta$, $\theta_4 = \theta- \pi$. Particles three and four are now taken to be outgoing, so momentum conservation gives
\be
p^1+p^2=p^3+p^4,
\ee
and the Mandelstam variables are defined by
\be
s= -(p_1+p_2)^2, \quad t= -(p_1-p_3)^2, \quad u = -(p_1-p_4)^2.  \label{mandelstamdefe}
\ee

The five polarization tensors for a massive spin-2 particle can be written in terms of vector polarizations in the following way:
\begin{subequations}
\begin{align}
\epsilon^{(\pm 2)}_{ \mu \nu} & = \epsilon^{(\pm 1)}_{ \mu} \epsilon^{( \pm 1) }_{\nu}, \\
\epsilon^{(\pm 1)}_{ \mu \nu} & = \frac{1}{\sqrt{2}} \left( \epsilon^{(\pm 1)}_{ \mu} \epsilon^{( 0) }_{\nu}+\epsilon^{(0)}_{ \mu} \epsilon^{( \pm 1) }_{\nu} \right), \\
\epsilon^{(0)}_{ \mu \nu} & = \frac{1}{\sqrt{6}} \left( \epsilon^{(1)}_{ \mu} \epsilon^{( -1) }_{\nu}+ \epsilon^{(-1)}_{ \mu} \epsilon^{( 1) }_{\nu}+2 \epsilon^{(0)}_{ \mu} \epsilon^{( 0) }_{\nu} \right).
\end{align}
\end{subequations}
To simplify the crossing equations we use vector polarizations corresponding to transversity states~\cite{Kotanski:1965zz, Kotanski:1970,deRham:2017zjm}, which for particle $j$ are given by
\begin{subequations} \label{eq:vecpolz}
\begin{align}
\epsilon^{(\pm 1)}_{ \mu}(p^j) & = \frac{i}{\sqrt{2} m} \left( p,E \sin \theta_j \pm i m\cos \theta_j,0,E \cos \theta_j \mp i m \sin \theta_j \right), \\
\epsilon^{(0)}_{ \mu}(p^j) & = \left( 0,0,1,0 \right).
\end{align}
\end{subequations}
The massive spin-2 propagator is given by
\be
P^{(m)}_{\mu_1 \mu_2, \nu_1 \nu_2} =-\frac{i}{2}\frac{  \Pi_{\mu_1 \nu_1} \Pi_{\mu_2 \nu_2} + \Pi_{\mu_1  \nu_2} \Pi_{\mu_2 \nu_1} -\frac{2}{3} \Pi_{\mu_1 \mu_2} \Pi_{\nu_1  \nu_2} }{p^2+m^2 - i\epsilon},
\ee
where
\be
\Pi_{\mu \nu} = \eta_{\mu \nu} + \frac{p_{\mu} p_{\nu}}{m^2},
\ee
and the massless spin-2 propagator in de Donder gauge is given by
\be
P^{(0)}_{\mu_1 \mu_2, \nu_1 \nu_2} =-\frac{i}{2} \frac{ \eta_{\mu_1 \nu_1} \eta_{\mu_2 \nu_2} + \eta_{\mu_1  \nu_2} \eta_{\mu_2 \nu_1} -\eta_{\mu_1 \mu_2} \eta_{\nu_1  \nu_2}}{p^2 - i\epsilon}.
\ee

\subsection{Constructing amplitudes}

We now briefly review how to use the on-shell vertices found in Section~\ref{sec:vertices} to construct the most general four-point amplitude with a given strong coupling scale. This procedure is discussed in greater detail in Ref.~\cite{Bonifacio:2018vzv}. 

Four-point tree amplitudes for massive spin-2 scattering will be denoted by $\mathcal{A}_{\tau_1 \tau_2 \tau_3 \tau_4}$, where $\tau_i$ labels the transversity of particle $i$. This can be written as the sum of contributions from exchange and contact diagrams,
\be \label{eq:amptotal}
\mathcal{A}_{\tau_1 \tau_2 \tau_3 \tau_4} = \mathcal{A}_{\tau_1 \tau_2 \tau_3 \tau_4}^{\rm exchange} +\mathcal{A}_{\tau_1 \tau_2 \tau_3 \tau_4}^{\rm contact}.
\ee
We calculate $\mathcal{A}_{\tau_1 \tau_2 \tau_3 \tau_4}^{\rm exchange}$ by combining the on-shell three-point vertices of Section~\ref{cubicsubsec} with massless and massive propagators and summing over channels,\footnote{The cubic vertices listed in Section~\ref{sec:vertices} are \textit{on-shell} invariant under interchanging the identical bosonic particles. This implies that the corresponding four-point exchange amplitude is only guaranteed to be Bose symmetric up to contact terms, since the internal leg is taken off shell. To ensure that the result is fully Bose symmetric, we first symmetrize in the two external legs of the cubic vertex when constructing exchange amplitudes, which is equivalent to adding certain contact terms.} as in Figure~\ref{fig:exchange}. This exchange amplitude is necessarily non-vanishing due to the contribution from graviton exchange through the compulsory minimal coupling interactions. At high energies it grows as some power of energy divided by some scale $\Lambda_k$ as defined in~\eqref{eq:cutoff}, so on its own this exchange amplitude would violate perturbative unitarity at energies around this scale. However, the total amplitude can have a higher strong coupling scale if there are cancellations between the high-energy parts of contact and exchange terms. For a given spectrum of particles, there is some maximum amount of cancellation that can occur without the entire amplitude vanishing and a corresponding upper bound on the cutoff.

To find out whether we can raise the strong coupling scale by some desired amount, we need to construct an ansatz for $\mathcal{A}_{\tau_1 \tau_2 \tau_3 \tau_4}^{\rm contact}$ that cancels the leading high-energy behavior of $\mathcal{A}_{\tau_1 \tau_2 \tau_3 \tau_4}^{\rm exchange}$. To do this we first strip off the kinematical singularities~\cite{Cohen-Tannoudji:1968lnm,Kotanski1968,Kotanski:1970}, so our contact ansatz takes the form
\be
\mathcal{A}^{\rm contact}_{\tau_1 \tau_2 \tau_3 \tau_4}(s,t) = \frac{a^{\rm contact}_{\tau_1 \tau_2 \tau_3 \tau_4}(s, t)+i \sqrt{s t u} \, b^{\rm contact}_{\tau_1 \tau_2 \tau_3 \tau_4}( s,t)}{\left(s-4m^2\right)^{ |\sum_i \tau_i|/2}},
\ee
where $a^{\rm contact}_{\tau_1 \tau_2 \tau_3 \tau_4}(s, t)$ and $b^{\rm contact}_{\tau_1 \tau_2 \tau_3 \tau_4}(s, t)$ are polynomials to be determined. The order of these polynomials is chosen such that $\mathcal{A}_{\tau_1 \tau_2 \tau_3 \tau_4}^{\rm contact}$ and $\mathcal{A}_{\tau_1 \tau_2 \tau_3 \tau_4}^{\rm exchange}$ are the same order in $E$ when expanded at high energies (with fixed scattering angle). Given some desired strong coupling scale, we choose the polynomial coefficients so that there are no terms suppressed by scales below this strong coupling scale in the high-energy expansion of the total amplitude, if this is possible. Next we impose crossing symmetry constraints on the contact terms by enforcing the following conditions \cite{Kotanski:1965zz, deRham:2017zjm},
\begin{align}
\mathcal{A}^{\rm contact}_{\tau_1 \tau_2 \tau_3 \tau_4}(s,t) & = e^{i\left(\pi- \chi_t\right) \sum_j \tau_j }\mathcal{A}^{\rm contact}_{-\tau_1 -\tau_3 -\tau_2 -\tau_4}(t,s),\\
\mathcal{A}^{\rm contact}_{\tau_1 \tau_2 \tau_3 \tau_4}(s,t) & = e^{i \left( \pi- \chi_u\right) \sum_j \tau_j }\mathcal{A}^{\rm contact}_{-\tau_1 -\tau_4 -\tau_3 -\tau_2}(u,t),
\end{align}
where
\begin{align}
e^{-i \chi_t} \equiv \frac{-st -2 i m \sqrt{s t u}}{\sqrt{ s(s-4m^2)t(t-4m^2)}}, \quad e^{-i \chi_u} \equiv \frac{-su +2 i m \sqrt{s t u}}{\sqrt{ s(s-4m^2)u(u-4m^2)}}.
\end{align}

At this point our ansatz for $\mathcal{A}^{\rm contact}_{\tau_1 \tau_2 \tau_3 \tau_4}$ is crossing symmetric, Lorentz invariant, and has the correct kinematical singularities, but it is not necessarily little group covariant. To ensure the ansatz has the correct little group transformations, we compare it to the general vertex $\mathcal{V}_{\rm contact}$ in Eq.~\eqref{eq:contactamp}, which is made from contractions of polarization tensors and hence transforms correctly under the little group.\footnote{As mentioned at the end of Section~\ref{sec:vertices}, the reason we do not deal directly with the vertex $\mathcal{V}_{\rm contact}$ is to avoid complications due to dimensionally-dependent identities.} We substitute four-dimensional kinematics in $\mathcal{V}_{\rm contact}$ and then require that $\mathcal{A}^{\rm contact}_{\tau_1 \tau_2 \tau_3 \tau_4}$ matches this for some choice of the polynomials $f_{I }(s,t)$. If this is possible, then the resulting contact term has all of the required kinematical properties and cancels the exchange terms to achieve the desired strong coupling scale. To find the maximum strong coupling scale we simply repeat this procedure with higher scales until it fails.  

\section{Results and examples}\label{sec:results}

We now present our result for a bound on the strong coupling scale of gravitationally coupled massive spin-2 particle and discuss several examples of theories subject to this result.   By carrying through the procedure described in Section \ref{sec:bound}, we find that the highest strong coupling scale in a parity-conserving, unitary EFT describing a massive spin-2 particle coupled to gravity is 
\be \Lambda_3=(m^2 M_P)^{1/3}.
\ee
Furthermore, if one of the particles has a wrong-sign kinetic term, then we find that the strong coupling scale can be raised all the way to $M_P$.  We know that these bounds are optimal, and not further lowered by amplitudes with other external particle configurations, because there are known theories that saturate them, which we now discuss.   

Although it is difficult to characterize the most general amplitude that achieves the $\Lambda_3$ strong coupling scale, we can find known theories among the optimal amplitudes by making further assumptions about the couplings.  For each example that we discuss with a known Lagrangian, we have explicitly calculated the four-point amplitude for massive spin-2 scattering from the associated Feynman rules and checked that it agrees with the amplitude we obtain from our procedure for raising the strong coupling scale, which provides a nontrivial check of our calculation.

\subsection{Ghost-free bigravity}

One example of a $\Lambda_3$ theory is Hassan-Rosen ghost-free bigravity \cite{Hassan:2011zd} (see Ref.~\cite{Schmidt-May:2015vnx} for a review), which is a generalization of general relativity to include interactions of the graviton with a massive spin-2 particle. The Lagrangian is given in terms of the two metrics $g_{\mu \nu}$ and $f_{\mu \nu}$,
\be
\mathcal{L} = \frac{M_g^2}{2} \sqrt{-g} R(g) +\frac{M_f^2}{2} \sqrt{-f} R(f)-\frac{m^2}{\frac{1}{M_g^2}+\frac{1}{M_f^2}} \sqrt{-g} \sum_{n=0}^4 \beta_n S_n \left({\mathbb X} \right), \label{bimetricactione}
\ee
where 
\be S_n^{\rm }({\mathbb X})= {\mathbb X}^{[\mu_1}_{\ \mu_1}{\mathbb X}^{\mu_2}_{\ \mu_2}\cdots {\mathbb X}^{\mu_n]}_{\ \mu_n}\ee
is the $n$-th elementary symmetric polynomial and 
\be {\mathbb X}^\mu_{\ \nu}\equiv \sqrt{g^{\mu \lambda} f_{\lambda \nu}}\ee
is defined by the matrix square root (which is unambiguous perturbatively).  The potential term is based on that of de Rham, Gabadadze, Tolley ghost-free massive gravity~\cite{deRham:2010kj} (see Refs.~\cite{Hinterbichler:2011tt,deRham:2014zqa} for reviews). There are two independent ``Planck masses" in this theory, $M_f$ and $M_g$, but the combination $\sqrt{M_f^2+M_g^2}$ sets the strength of the gravitational interactions and so this corresponds to what we have been calling $M_P$. The strong coupling scale is set by the smaller of $M_f$ and $M_g$,
\be
\Lambda = \left( m^2 \min(M_f, M_g) \right)^{1/3}.
\ee
This is of order $\Lambda_3 = \left( m^2 M_P \right)^{1/3}$ for $M_f \approx M_g$, when it is at its largest, but can be made parametrically smaller, e.g. by sending $M_f/M_g \rightarrow 0$.

Demanding flat spacetime as a solution and normalizing so that $m$ in \eqref{bimetricactione} is the mass of the massive particle, the parameters $\beta_i$ in \eqref{bimetricactione} can be given in terms of two free parameters.  We choose these as $c_3$ and $d_5$ given by
\begin{subequations}
\begin{align}
\beta_0 & = 48 d_5+ 24 c_3  -6, \\
\beta_1 & = - 48 d_5  - 18 c_3 +3, \\
\beta_2 & = 48 d_5+ 12 c_3 -1, \\
\beta_3 & =  - 48 d_5-6 c_3, \\
\beta_4 & = 48 d_5.
\end{align}
\end{subequations}
Around flat spacetime the mass eigenstates are related to the metrics by
\begin{align}
g_{\mu \nu} = \eta_{\mu \nu} + \frac{2}{M_g} \frac{M_g u_{\mu \nu} +M_f v_{\mu \nu}}{\sqrt{M_f^2+M_g^2}}, \quad f_{\mu \nu} = \eta_{\mu \nu} + \frac{2}{M_f} \frac{M_f u_{\mu \nu} -M_g v_{\mu \nu}}{\sqrt{M_f^2+M_g^2}}, \label{eq:mastend}
\end{align}
where $u_{\mu \nu}$ describes a canonically normalized massless spin-2 particle and $v_{\mu \nu}$ a canonically normalized spin-2 particle with mass $m$. 

In this theory, the nonzero cubic couplings $a_i$ of \eqref{cubicseqpa} are given by
\be \label{eq:bgcubicsa}
a_3=2a_2=\frac{4(M_f^2-M_g^2)}{M_f M_g \sqrt{M_f^2+M_g^2}}, \quad a_1 =\frac{3(1-4 c_3)\sqrt{M_f^2+M_g^2}}{M_f M_g},
\ee
and the nonzero couplings $b_i$ of \eqref{eq:bvertex} are given by
\be \label{eq:bgcubicsb}
2b_1 = b_2 = 2b_3= \frac{4}{\sqrt{M_f^2+M_g^2}}.
\ee
By setting the cubic couplings to these values and restricting to $\Lambda_3$ amplitudes in our procedure for raising the strong coupling scale, we recover precisely the four-point amplitude of ghost-free bigravity.
As an aside, this implies that the ghost-free bigravity amplitude with $c_3=1/4$ is the unique parity-even $\Lambda_3$ amplitude without a Shapiro time advance, since the absence of time advances constrains the cubic couplings to satisfy $a_3=2a_2$, $a_1=0$ and $2b_1 = b_2 = 2b_3$~\cite{Bonifacio:2017nnt}.

\subsection{Quadratic curvature gravity}

From our procedure we find that there is an amplitude with a strong coupling scale above $\Lambda_3$ if we permit \textit{imaginary} cubic couplings with the following relative values:
\be \label{eq:QCGcouplings}
4 b_1 = 2 b_2= 4 b_3  = \pm 2 i a_2 = \pm i a_3.
\ee 
Such imaginary couplings correspond to a real Lagrangian in which one field has a wrong-sign kinetic term. The resulting amplitudes become strongly coupled at the Planck scale, $M_P$, so this is the maximum possible strong coupling scale with two spin-2 fields when one of them is a ghost.

In fact, there does exist a well-known
theory of a ghostly massive spin-2 particle coupled to gravity, namely quadratic curvature gravity~\cite{Stelle:1977ry}. The action of this theory is
\be
S = M_P^2 \int d^4 x\; \sqrt{-g} \left( \frac{1}{2} R + \frac{1}{4 m^2} C_{\mu \nu \rho \sigma} C^{\mu \nu \rho \sigma} \right),
\ee
where $C_{\mu \nu \rho \sigma} $ is the Weyl tensor. This can be written in second-order form using a symmetric auxiliary tensor field $v_{\mu \nu}$,
\be
S = M_P^2  \int d^4 x\; \sqrt{-g} \left( \frac{1}{2} R + v^{\mu \nu} G_{\mu \nu} -\frac{1}{2} m^2 \left( v_{\mu \nu} v^{\mu \nu} - v^2 \right) \right),
\ee
where indices are contracted with $g_{\mu \nu}$.
If we expand the metric around a flat background as
\be
g_{\mu \nu} = \eta_{\mu \nu} + 2 u_{\mu \nu}+ 2 v_{\mu \nu}
\ee
and rescale $u_{\mu \nu} \rightarrow u_{\mu \nu}/M_P$, $v_{\mu \nu} \rightarrow v_{\mu \nu}/M_P$, then the resulting kinetic terms shows that $u_{\mu \nu}$ is a healthy massless spin-2 field and $v_{\mu \nu}$ is a ghostly massive spin-2 field with mass $m$. We can make the kinetic term for $v_{\mu \nu}$ healthy at the expense of generating imaginary cubic couplings by sending $v_{\mu \nu} \rightarrow i v_{\mu \nu}$ and $m \rightarrow i m$.\footnote{Alternatively, we can have real cubic couplings with a healthy massive field and a ghostly massless field by flipping the overall sign of the action. The resulting amplitudes are the same up to an overall sign.} After this we find that the massless cubic vertex is the same as in Einstein gravity and the other nonzero cubic couplings are given by
\be \label{eq:cubicsQCG}
4 b_1= 2 b_2=4 b_3 = -2ia_2=-ia_3 =\frac{8}{M_P}.
\ee
This corresponds precisely to one of the solutions in Eq.~\eqref{eq:QCGcouplings} after imposing the gauge invariance conditions~\eqref{eq:bconstraints}. The other solution comes from sending $a_i \rightarrow - a_i$, which corresponds to flipping the sign of the massive field and leaves this four-point amplitude invariant. 

We thus deduce that any theory of a massive and massless spin-2 field with a strong coupling scale above $\Lambda_3$ contains a ghost and has the same four-point massive spin-2 amplitudes as quadratic curvature gravity. The leading helicity amplitudes, which at high energy are of the form $E^2/M_P^2$, come from polarization configurations of the form $TTTT$, $TTVV$, and $TTSS$, where $S$, $V$, and $T$ stand for the scalar, vector, and tensor helicity polarizations of the massive spin-2 particle. The four-point amplitudes in quadratic curvature gravity with external massless gravitons are known to coincide with those of Einstein gravity~\cite{Dona:2015tra}, so these amplitudes also become of order one around $M_P$.  This result is somewhat puzzling given the apparent renormalizability and asymptotic freedom of quadratic curvature gravity \cite{Stelle:1977ry,Fradkin:1981hx, Avramidi:1985ki}.\footnote{In v2 of Ref.~\cite{Hinterbichler:2015soa}, a St\"uckelberg analysis of quadratic curvature gravity was used to argue that the high-energy amplitudes have no strong coupling scale.   The last step of that analysis, used to get from the scale $M_P$ to $\infty$, involved a hyperbolic field redefinition in the Lorentzian field space followed by a limit to infinity along a non-compact direction in this field space which kills off the interactions.  This limit is actually not justified because it is not an invertible redefinition (the interactions that were killed off cannot be recovered by undoing the field redefinition).  Without this final step, the strong coupling scale of that analysis is indeed $M_P$, consistent with our findings here.}

It is interesting to compare the cubic couplings in ghost-free bigravity, \eqref{eq:bgcubicsa} and \eqref{eq:bgcubicsb}, to those in quadratic curvature gravity, \eqref{eq:cubicsQCG}. We see that these couplings agree if we take the scaling limit
\be \label{eq:scalinglimit}
M_g \rightarrow \infty, \quad M_f^2+M_g^2 \equiv M_P^2 \quad \text{fixed}\, ,
\ee
which sends $M_f \rightarrow \pm i M_g$.  In fact, we find that in this limit the massive four-point amplitudes of these two theories agree exactly, so the amplitude of quadratic curvature gravity is contained in the amplitude of ghost-free bigravity. This relationship follows from the results of Ref.~\cite{Paulos:2012xe}, where it was shown that the Lagrangian of ghost-free bigravity reduces to that of quadratic curvature gravity in the scaling limit~\eqref{eq:scalinglimit}. Another correspondence between ghost-free bigravity and higher-derivative theories was discussed in Ref.~\cite{Hassan:2013pca}.

\subsection{Pseudo-linear theory}

So far we have considered theories with long-range gravity.  If we relax this assumption by turning off the Einstein-Hilbert vertex \eqref{eq:GRvertex}, then we get theories with only a linear spin-2 gauge invariance and there are no requisite minimal coupling interactions.  We then have a choice for what scales we give to interactions and hence there is no invariant meaning to finding the highest strong coupling scale. However, a comparable property we can study is how much the amplitude grows with energy for $E \gg m$, as explained in Ref.~\cite{Bonifacio:2018vzv}. A strong coupling scale of $\Lambda_3$ corresponds to four-point amplitudes that grow with energy like $E^6$, so amplitudes with this growth are the analogue of $\Lambda_3$ theories. 

Allowing for any values of the couplings $b_i$ in \eqref{eq:bvertex}, we find no additional non-vanishing\footnote{When the only non-vanishing cubic coupling is $b_6$, the exchange amplitude is analytic and can be cancelled entirely by a contact term.} amplitudes that grow slower than $E^6$, regardless of the signs of the kinetic terms.  But there are new $E^6$ amplitudes that are different from those allowed with an Einstein-Hilbert structure.  One of these corresponds to the ghost-free pseudo-linear theory containing a massive spin-2 particle and a massless spin-2 particle with a linear gauge symmetry~\cite{Bonifacio:2018van}. The Lagrangian of this theory is given by
\be
\mathcal{L} = \frac{1}{2}  u^{\mu \nu} (\mathcal{E} u)_{\mu \nu}+\frac{1}{2}   v^{\mu \nu} (\mathcal{E} v)_{\mu \nu}- \frac{m^2}{2}\left(v^{\mu \nu} v_{\mu \nu} -v^2 \right) + \frac{\lambda}{2 M_P} \epsilon^{\mu_1 \mu_2 \mu_3 \mu_4} \epsilon^{\nu_1 \nu_2 \nu_3 \nu_4} v_{\mu_1 \nu_2} v_{\mu_2 \nu_2} \partial_{\mu_3}\partial_{\nu_3} u_{\mu_4 \nu_4} ,\label{eq:plbigravity}
\ee
where 
\be
(\mathcal{E} u)_{\mu \nu} \equiv \Box u_{\mu \nu} - 2 \partial_{( \mu} \partial^{\lambda} u_{\nu) \lambda}+\partial_{\mu} \partial_{\nu} u + \eta_{\mu \nu} \left( \partial^{\lambda} \partial^{\rho} u_{\lambda \rho} - \Box u \right)
\ee
defines the standard spin-2 linear kinetic term and traces are taken with $\eta^{\mu \nu}$. The corresponding on-shell cubic vertex $\mathcal{V}_b$ has $b_3 = \lambda/M_P$, with the other $b_i$ equal to zero. 
By restricting the cubic couplings to take these values in our procedure for raising the strong coupling scale, we obtain the amplitude generated by~\eqref{eq:plbigravity}.  There are also other possible $E^6$ interactions containing only the massive field that we have not written here \cite{Folkerts:2011ev,Hinterbichler:2013eza}, which would give non-zero values for some of the $a_i$'s in \eqref{cubicseqpa}.

\subsection{Gravitationally coupled pseudotensor}

We have so far assumed that the massive spin-2 particle transforms as a tensor under parity. However, consistent with our assumption of parity conservation, we can also consider a massive particle that transforms as a pseudotensor under parity. Such a particle would still interact with Einstein gravity through the parity-even vertices in Eq.~\eqref{eq:bvertex}, but it would have parity-odd self-interactions involving an epsilon symbol, rather than the parity-even interactions of \eqref{cubicseqpa}.  Repeating our calculation with parity-odd self-interactions, we find that $\Lambda_3$ is still the highest strong coupling scale.  We also find $\Lambda_3$ amplitudes with correct-sign kinetic terms that describe a gravitationally coupled pseudotensor with self-interactions described by the following two-derivative parity-odd vertex \cite{Bonifacio:2018vzv}:
\be
\tilde{\mathcal{V}}= i \frac{ 4 \gamma}{M_P} \left( z_{13} z_{23} \varepsilon (p_{1} p_2 z_{1} z_2 )-z_{12} z_{23} \varepsilon (p_{1} p_2 z_{1} z_3)+z_{12} z_{13} \varepsilon (p_{1} p_2 z_{2} z_3) \right),
\ee
where $\varepsilon ( \cdot )$ denotes the contraction of the antisymmetric symbol with the enclosed vectors in the order shown.
The other cubic couplings are given by
\be
2b_1= b_2= \frac{4}{M_P}, \quad b_3 = \frac{2 ( \gamma^2+1) }{M_P}, 
\ee
where $\gamma$ is a free parameter. 

There is no known ghost-free theory describing a gravitationally coupled massive pseudotensor, but the existence of this amplitude is suggestive. When $\gamma=0$, the amplitude reduces to that of the special bigravity theory with $c_3 =1/4$ and $M_f=M_g$, which has a $\mathbb{Z}_2$ symmetry under interchanging $f_{\mu \nu}$ and $g_{\mu \nu}$. This symmetry implies that the massive tensor $v_{\mu\nu}$ as defined in \eqref{eq:mastend} has no preferred parity, since only even powers of this field can appear. The hypothetical parity-odd theory would therefore be a deformation of this special theory that differs from the usual bigravity theory. However, by calculating amplitudes from a general gauge-invariant two-derivative Lagrangian, we find that the four-point amplitude with one external graviton and three massive legs becomes strong by the lower scale $\Lambda_{7/2} = (m^5 M_p^2)^{1/7}$ when $\gamma \neq 0$, assuming that the massive amplitude becomes strong at $\Lambda_3$.
 
\section{Discussion} \label{sec:discussion}

We have explored how the existence of Einstein gravity can constrain theories of massive higher-spin particles. We found that the strong coupling scale in a parity conserving theory of a gravitationally coupled massive spin-2 particle cannot exceed $\Lambda_3 = \left(m^2 M_P \right)^{1/3}$.  We have assumed only standard properties such as unitary, locality, Lorentz invariance, and crossing symmetry. We have made no additional assumptions about properties of the Lagrangian, except that the number of derivatives is bounded (but still arbitrary).  If we allow for wrong-sign kinetic terms, then the strong coupling scale can be raised to $M_P$, as realized by quadratic curvature gravity.

One consequence of this result is that
there is a lower bound on the mass of any isolated and weakly coupled neutral massive spin-2 particle in our universe, assuming gravity is described by general relativity.  For example, if there are no other new particles or strong coupling effects below the neutrino mass scale, then we must have $\Lambda_3 \gtrsim 10^{-3} \, \rm{eV}$. This then gives a conservative lower bound on the spin-2 mass of $m \gtrsim 10^{-18} \, \rm{eV}$. Any neutral massive spin-2 particle that is lighter than this must either be accompanied by other new particles or become strongly coupled below the neutrino mass scale. 
We can also apply this result to constrain the spectra of large-$N$ confining gauge theories. Suppose there is a confining large-$N$ theory coupled to gravity whose lightest state is a spin-2 meson or glueball with mass $m$.\footnote{The intuitive picture of mesons as quarks joined by a string of flux suggests that the lowest energy state should not have angular momentum and hence would be a scalar, though we are not aware of any proof of this.} Since the theory is weakly coupled, our result implies that the mass of the next lightest state must be less than $(m^2 M_P)^{1/3}$, so the gap between these states cannot be arbitrarily large.

We have focused in this paper on the case of a massive spin-2 particle coupled to gravity, but it is interesting to speculate about the generalization to higher spins.  Rahman has conjectured in Ref.~\cite{Rahman:2009zz} that the maximum strong coupling scale of a massive spin-$s$ particle coupled only to Einstein gravity is given by 
\be \Lambda_{2s-1} \equiv \left( m^{2s-2} M_P \right)^{1/(2s-1)}. \label{eq:conjecture}\ee
We give now some additional motivation for this conjecture by using the weak gravity conjecture. 
Suppose the low-energy EFT of a theory of quantum gravity is described by a single charged massive spin-$s$ particle with $s > 1$. Then by the result of Ref.~\cite{Porrati:2008ha}, this EFT violates tree-level unitarity at some scale $\Lambda$, where
\be \label{eq:EMbound}
\Lambda \leq \frac{m}{q^{1/(2s-1)}}.
\ee
This strong coupling scale is nonzero in the limit $q \rightarrow 0$ with $m$ fixed, which runs afoul of the folk theorem that there are no global symmetries allowed in quantum gravity. One way to resolve this is if the lightest state satisfies the bound on the charge given by the weak gravity conjecture~\cite{ArkaniHamed:2006dz},
\be \label{eq:QGC}
q \geq \frac{m}{M_P},
\ee
which means we cannot send $q\rightarrow 0$ with $m$ fixed. Combining these inequalities then gives
\be \label{eq:conjecture2}
\Lambda \leq \Lambda_{2s-1} \equiv \left( m^{2s-2} M_P \right)^{1/(2s-1)},
\ee
so now we have $\Lambda \rightarrow 0$ as $q \rightarrow 0$.  This would imply that the maximum strong coupling scale of the gauge theory is less than the maximum strong coupling scale associated with gravity if and only if Eq.~\eqref{eq:QGC} is satisfied, which is similar to the findings of Ref.~\cite{Heidenreich:2017sim}.  

For $s=2$, the conjecture \eqref{eq:conjecture} reduces to the bound found in this paper.  It is shown to be true for $s=3/2$ in Ref.~\cite{Rahman:2011ik}, and we have checked it for $s=1$ by confirming that the maximum strong coupling scale for the massive spin-$1$ four-point amplitude with graviton exchange is $M_P$ \cite{Chowdhury:2018nfv}.
It would be interesting to find a proof or counterexample when $s > 2$.

{\bf Acknowledgements:}  We would like to thank Clifford Cheung, Juan Maldacena, Rakibur Rahman and Rachel Rosen for helpful conversations and correspondence.

\appendix

\section{Constraints from gauge invariance} \label{app:gauge}

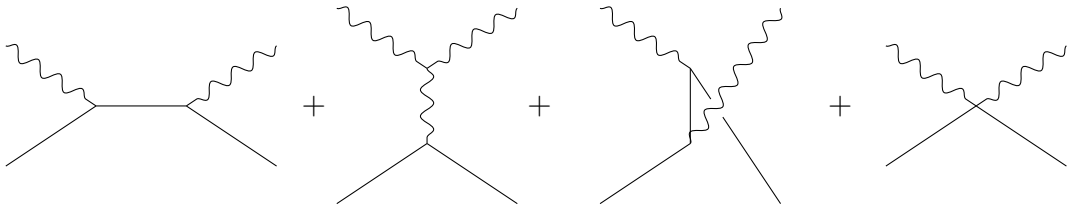
\begin{figure}[!ht]
\begin{center}
\begin{tikzpicture}[node distance=.8cm and 1.2cm]
\coordinate (s1);
\coordinate[below right=of s1] (svertex1);
\coordinate[below left=of svertex1] (s2);
\coordinate[right=1.2cm of svertex1] (svertex2);
\coordinate[above right=of svertex2] (s3);
\coordinate[below right=of svertex2] (s4);

\draw[graviton] (s1) -- (svertex1);
\draw[mgraviton] (svertex1) -- (s2);
\draw[graviton] (s3) -- (svertex2);
\draw[mgraviton] (svertex2) -- (s4);
\draw[mgraviton] (svertex1) -- (svertex2);

\coordinate[right=2cm of svertex2,label=left:$+$] (mid1);
\coordinate[right=1.2cm of mid1] (tmid);

\coordinate[above=.5cm of tmid] (tvertex1);
\coordinate[above left=of tvertex1] (t1);
\coordinate[above right=of tvertex1] (t3);
\coordinate[below=1cm of tvertex1] (tvertex2);
\coordinate[below left=of tvertex2] (t2);
\coordinate[below right=of tvertex2] (t4);

\draw[graviton] (t1) -- (tvertex1);
\draw[graviton] (t3) -- (tvertex1);
\draw[graviton] (tvertex1) -- (tvertex2);
\draw[mgraviton] (t2) -- (tvertex2);
\draw[mgraviton] (t4) -- (tvertex2);

\coordinate[right=3cm of mid1,label=left:$+$] (mid2);

\coordinate[right=3.5cm of tvertex1] (uvertex1);
\coordinate[above left=of uvertex1] (u1);
\coordinate[above right=of uvertex1] (u3);
\coordinate[below=1cm of uvertex1] (uvertex2);
\coordinate[below left=of uvertex2] (u2);
\coordinate[below right=of uvertex2] (u4);

\draw[mgraviton] (uvertex1) -- (uvertex2);
\draw[mgraviton] (u2) -- (uvertex2);
\draw[mgraviton] (u4) -- (uvertex1);
\draw [white, line width=6] (9.37,-.71) -- (9.53,-.95);
\draw[graviton] (u1) -- (uvertex1);
\draw[graviton] (u3) -- (uvertex2);

\coordinate[right=4cm of mid2,label=left:$+$] (mid3);

\coordinate[right=1.5cm of mid3] (contactvertex);
\coordinate[above left=of contactvertex] (c1);
\coordinate[above right=of contactvertex] (c3);
\coordinate[below left=of contactvertex] (c2);
\coordinate[below right=of contactvertex] (c4);

\draw[graviton] (c1) -- (contactvertex);
\draw[graviton] (c3) -- (contactvertex);
\draw[mgraviton] (c2) -- (contactvertex);
\draw[mgraviton] (c4) -- (contactvertex);

\end{tikzpicture}
\end{center}
\caption{\small Diagrams contributing to graviton (wavy line) Compton scattering with a massive spin-2 particle (solid line). For simplicity we ignore the vertex with one massive and two massless particles.}
\label{fig:gravicompton}
 \end{figure}

Here we discuss the origin of the constraints \eqref{eq:bconstraints} on the cubic couplings between two massive particles and one massless particle.
These constraints are satisfied automatically in a theory with a generally covariant Lagrangian, but here we want to see them using $S$-matrix arguments. Assuming the presence of the GR cubic vertex \eqref{eq:GRvertex}, we consider graviton Compton scattering off a massive spin-2 particle, as depicted in Figure~\ref{fig:gravicompton}.  Gauge invariance requires that the overall four-point amplitude is invariant under the replacement $z_i \rightarrow z_i + \xi p_i$ for the external massless particles. 

We construct this amplitude by combining on-shell cubic vertices with propagators to make the exchange diagrams and then adding a general contact vertex.  The sum of the exchange diagrams will not in general be gauge invariant, so its gauge variation must be cancelled by the contact vertex.
For the contact vertex, we need to consider terms that are symmetric under interchanging the identical particles and that have up to fourteen derivatives, since these are the only terms whose gauge variation can help cancel the gauge variation of the exchange terms.\footnote{This is not strictly true since the number of derivatives can decrease under a gauge variation when there are massive particles. Terms with even more derivatives might contribute lower-order pieces that help cancel the gauge variation.  We do not consider this possibility, which represents a possible loophole in the analysis of this appendix.} There are $661$ different contact terms of this form.

Imposing gauge invariance while allowing for a general local contact vertex to cancel the gauge variation of the exchange diagrams gives constraints on the cubic couplings similar to those arising from the spinor helicity four-particle test~\cite{Benincasa:2007xk}.  Whereas gauge invariance of the four-point amplitude is manifest in spinor-helicity variables and imposing locality gives constraints~\cite{Arkani-Hamed:2017jhn}, here locality is manifest and imposing gauge invariance gives the constraints.  An explicit calculation shows that the corresponding Ward identities can only be satisfied for non-vanishing $b_i$ in \eqref{eq:bvertex} if we have
\be \label{eq:gaugecons2}
2 b_1 = b_2 =  \frac{4}{M_P}.
\ee
The two-derivative vertex $\mathcal{B}_3$ is quadratic in the graviton momentum and it can be freely adjusted by adding non-minimal interactions containing the Riemann tensor, which explains why $b_3$ is unconstrained by gauge invariance.

As we have ignored dimensionally-dependent identities and parity-odd interactions, this argument is only rigorously true for $D>7$. 
However, we expect the conclusion to be fairly universal since these couplings also follow from the usual minimal coupling procedure in the Lagrangian formulation. Indeed, the constraint on $b_1$ is a consequence of the $S$-matrix equivalence principle~\cite{Weinberg:1964ew}, which holds at least for $D \geq 4$. But in principle there could also be exotic theories that exist by taking advantage of degenerate kinematics in low dimensions. For simplicity, we have also ignored in this discussion the higher-derivative graviton cubic interactions and cubic vertices with two gravitons and one massive spin-2 particle, but including these does not affect the conclusion.

\bibliographystyle{utphys}
\addcontentsline{toc}{section}{References}
\bibliography{bigravity-paper-arxiv2}

\end{document}